\documentclass{emulateapj}
\usepackage{psfig,times,natbib,amssymb}

\newcommand{\chandra}{{\em Chandra}}

\newcommand\msun{{$M_{\odot}$}}
\newcommand{\kms}{{km\,s$^{-1}$}}

\newcommand{\adspr}{{Advances in Space Research}}

\begin{document}

\title{The Kinematics of Kepler's Supernova Remnant as Revealed by Chandra}


\author{Jacco Vink\altaffilmark{1} }

\email {j.vink@astro.uu.nl}
\altaffiltext{1}{Astronomical Institute, University Utrecht, P.O. Box 80000, 3508TA Utrecht, The Netherlands}

\begin{abstract}
I have determined the expansion of the supernova remnant
of SN1604 (Kepler's supernova) 
based on archival \chandra\ ACIS-S observations made in
2000 and 2006.
The measurements were done in several distinct energy bands, and
were made for the remnant as a whole, and for six individual
sectors.
The average expansion parameter indicates that the remnant expands
on average as $r \propto t^{0.5}$, but there are significant differences in
different parts of the remnant:
the bright northwestern part expands as $r \propto t^{0.35}$,
whereas the rest of the remnant's expansion shows an expansion
$r \propto t^{0.6}$. The latter is consistent with an explosion in which the
outer part of the ejecta has a negative power law slope for density 
($\rho \propto v^{-n}$) of $n=7$, or with an exponential density 
profile($\rho \propto \exp(-v/v_e)$). The expansion parameter in the
southern region, in conjunction with the shock radius, indicate a rather
low value ($<5 \times 10^{50}$~erg)
for the explosion energy of SN1604 for a distance of 4~kpc.
An higher explosion energy is consistent with the results, 
if the distance is larger.

The filament in the eastern part of the remnant, which is dominated by
X-ray synchrotron radiation seems to mark a region with  a fast shock
speed $r \propto t^{0.7}$, corresponding to a shock velocity of
$v= 4200$~\kms, for a distance to SN1604 of 4 kpc. This is consistent
with the idea that X-ray synchrotron emission requires shock velocities
in excess of $\sim 2000$~\kms.

The X-ray based expansion measurements reported are consistent with
results based on optical and radio measurements,
but disagree with previous X-ray measurements based on ROSAT and Einstein
observations.

\end{abstract}

\keywords{X-rays: observations individual (Kepler/SN 1604) --supernova remnants }

\section{Introduction}
Among the more than 250 known Galactic supernova remnants, the remnants
of the historical supernovae hold a special place \citep{stephenson02}.
This has partially to do with the fascination for historical events
that caught the imagination of astronomers in ancient China,
the Middle East, and in renaissance Europe. 
But also more scientific reasons make the study of 
historical supernova remnants worthwhile: we know the exact
age of the objects. Moreover, the historical supernova remnants are
among the youngest supernova remnants, which means that their X-ray emission
is largely dominated by shock heated ejecta rather than shocked 
interstellar matter. Historical remnants are therefore
prone to offer new insights into the supernova explosion properties.

The youngest historical supernova remnant (SNR) 
is SN1604 \citep{stephenson02}, 
also known Kepler's SNR (Kepler for short).\footnote{
Cas A is a younger remnant, but not strictly an historical remnant, because 
the supernova was likely not observed \citep{stephenson02}.}
The supernova was first sighted on the evening of October 9, 1604, 
low above the horizon. It owes its early discovery probably to the 
much anticipated simultaneous
conjunction of Jupiter, Saturn and Mars.
Johannes Kepler lived in Prague at that time, and suffered from bad weather.
However, from the first reports on he took a keen interest in the new star,
and after the weather improved, he  started his own observations. 
The results of his
observations and his correspondence with other observers led to his book
on the supernova, ``De Stella Nova'', which was published in 1606.

The SNR of SN1604 has been a puzzling object for some time
\citep{blair05}.
Both the historical light curve of the supernova \citep{baade43} and 
its relatively 
high Galactic latitude ($l=4.5$\degr, $b=6.4$\degr) suggest that
Kepler is the
result of a Type Ia supernova. However, optical observations of the
remnant reveal the
presence of copious amounts of nitrogen, in particular in the Northwest.
Nitrogen is an element associated with stellar winds rather than Type Ia 
supernovae. This prompted \citet{bandiera87} to suggest that the progenitor
was a massive, runaway star, thus explaining both the origin
of nitrogen and the high Galactic latitude of the supernova. Nevertheless,
the X-ray spectrum of Kepler indicates  a large abundance of iron,
which points to a Type Ia supernova \citep{kinugasa99,cassam04}.
A recent deep \chandra\ observation revealed no evidence for a neutron
star \citep{reynolds07}, which would be expected
if SN1604 was a core collapse supernova.
Having considered all evidence, \citet{reynolds07} conclude that Kepler is
the remnant of a Type Ia supernova from a relatively massive progenitor star,
attributing the nitrogen to stellar wind loss from either the white dwarf 
progenitor, or the companion star.

The situation concerning the kinematics of Kepler is
equally confusing.
\citet{dickel88} studied the expansion of Kepler in the radio, based on
VLA 6 cm and 20 cm data covering a time span of 4 yr. 
They found expansion rates indicating significant
deceleration with, on average, an expansion rate consistent with a
radial expansion law
$r \propto t^{0.5}$, and in the northern part even as low as  
$r \propto t^{0.35}$.
Optical expansion measurements, 
based on ground based and Hubble Space Telescope imaging of the bright 
H$\alpha$ filaments in the Northwest covering a time span of 16.33 yr,
showed proper motions of 1.3\arcsec\ to 1.6\arcsec\ \citep{sankrit05}. 
This corresponds to an expansion  law following  $r \propto t^{0.35}$, 
consistent with the radio measurements.
However, X-ray expansion measurements, based on Einstein and ROSAT observations
with the high resolution imagers onboard these two satellites indicated
nearly free expansion: $r \propto t^{0.93}$ \citep{hughes99}.
Note that of all wavelength regimes the X-ray emission is most closely
associated with the dynamics of the remnant, since the shock heated
plasma has most of its emission in X-rays, and best compares with
hydrodynamic simulations. The optical emission is confined to
a region close to the shock front, whereas the kinematics as derived from
the  radio emission should be related to the X-ray emitting
plasma, but the mismatch between radio and X-ray expansion
measurements for Kepler, but also Cas A \citep{vink98a,koralesky98,delaney03},
and Tycho's SNR \citep{hughes00c} complicates the interpretation.

\begin{table}
\caption{Observations \label{table-obs}}
\begin{tabular}{llll}\noalign{\smallskip}\hline\noalign{\smallskip}
Observation & Start date & MJD & Exposure (ks) \\

\noalign{\smallskip}\hline\noalign{\smallskip}
116 &	2000-06-30T12:04:56 &	51725.5034 &	49.45\\
\noalign{\smallskip}
6714&	2006-04-27T23:13:49 & 53852.97 &	159.84 \\
6715&	2006-08-03T16:52:09 & 53950.70 &	161.17 \\
6716&	2006-05-05T19:18:18 & 53860.80 &	160.05 \\
6717&	2006-07-13T19:04:25 & 53929.79 &	108.18 \\
6718&	2006-07-21T14:55:14 & 53937.62 &	109.18 \\
7366&	2006-07-16T05:24:07 & 53932.23 &	52.12 \\

\noalign{\smallskip}\hline
\end{tabular}
\end{table}

Resolving the discrepancy between radio and X-ray expansion
measurements, and obtaining the overall kinematics of Kepler's SNR
is important for several reasons:
First of all the distance to the SNR is poorly known. One
way to estimate the distance is to measure both the proper motion and the shock
velocity. The latter can be done independent of the proper motion by measuring
the broad  H$\alpha$ emission line, which is the result of charge transfer from
neutral hydrogen entering the shock to shock heated protons. The broadening
of the line therefore is a measure of the proton temperature, which is linked
to the shock velocity. For Kepler this method yields a shock velocity 
of 1550-2000 \kms \citep{fesen89,blair91}, 
implying a distance of $3.9^{+1.4}_{-0.9}$~kpc \citep{sankrit05}. 
This is slightly lower, but consistent with
the most recent measurement based on HI absorption features, 
$4.8\pm 1.4$~kpc \citep{reynoso99}.

Another issue is the dynamical state of the SNR.
As long as the mass of the ejecta dominates over the the mass of shock
heated circumstellar material the remnant is little decelerated, and said
to be in the free expansion phase with $R\propto t$ 
\citep[see, however][]{truelove99}. Once the energy in the shock heated
circumstellar medium dominates the total energy, the SNR  
is said to have entered
the Sedov-Taylor stage of its evolution with $R\propto t^{0.4}$ for a
uniform density medium.
If SN1604 is a Type Ia supernova one expects it to have a relatively
low ejecta mass of 1.4\msun. A high expansion rate, as found in X-rays,
is therefore puzzling.

Finally, in recent years it has been found that all young SNRs have thin
filaments at the shock front, whose emission is dominated by
synchrotron radiation \citep[e.g.][]{vink03a,bamba05,voelk05}.
The widths of these filaments can be used to infer
magnetic fields  \citep[e.g.]{vink03a,bamba05,voelk05,warren05}, which
turn out to be relatively high. This is therefore evidence for
cosmic ray driven magnetic field amplification \citep{bell01,bell04}.
However, it is not quite clear how the magnetic field scales with
density and shock velocity; it could be either $B^2 \propto \rho v^2$
\citep{voelk05} or  $B^2 \propto \rho v^3$ \citep{bell04,vink06b}.
The range in densities among the SNRs is quite large (a factor of 100
from SN1006 to Cas A), so the
dependency of $B^2$ on $\rho$ can be determined reasonably well. 
However,
X-ray synchrotron filaments only arise for high shock velocities
$v \gtrsim  2000$~\kms \citep{aharonian99}, and since all
known SNRs have $v \lesssim 6000$~\kms, 
the dynamic range in velocity is not very high, as compared to
the dynamic range in densities.
Moreover, the uncertainties in the measured 
velocities is quite high. A more accurate
assessment of the shock velocities in
those regions where X-ray synchrotron filaments have been
found \citep{reynolds07} is therefore valuable. 
In addition,
the cut-off photon energy of X-ray synchrotron radiation
depends not only on 
$v_s$, but also on the cosmic ray diffusion parameter \citep{aharonian99}:
\begin{equation}
E_{\rm cut-off} = 0.5 \eta^{-1} \bigl( \frac{v_s}{2000\, {\rm km\, s^{-1}}} \bigr)^2 {\rm keV},
\end{equation}
with $\eta$  a parameter that relates the actual diffusion coefficient
to the diffusion coefficient for the most optimal diffusion coefficient for 
fast cosmic
ray acceleration, the so called Bohm diffusion. For Bohm diffusion $\eta=1$,
and the magnetic field is highly turbulent \citep[see][for a review]{malkov01}.
Apparently, $\eta\approx 1$ for young SNRs 
\citep{vink04b,vink06b,stage06}. 
Accurate shock velocity measurement are therefore important for
estimating the diffusion constant, and the related turbulence
of the magnetic field.

Here I present expansion measurements based
on archival \chandra\ data obtained in 2000 and 2006.

\begin{figure}
\includegraphics[angle=-90,width=\columnwidth]{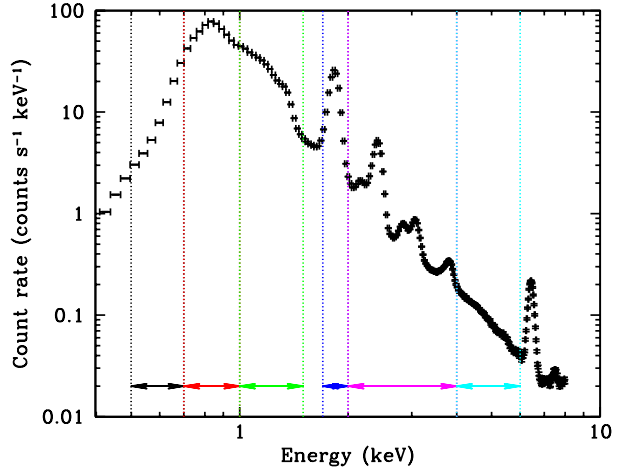}
\caption{
The \chandra\ ACIS-S X-ray spectrum of the entire remnant, based on
all 2006 observations \citep{reynolds07}. The arrows and dotted
lines mark the energy bands used to measure the expansion.
\label{fig-spectrum}
}
\end{figure}

\section{Data and Method}

\subsection{Observations}
\chandra\ observed Kepler several times, the first time
in June 2000 as part of the Guaranteed Time Observation program.
The most recent observation was made in 2006 as part of the
Large Program \citep{reynolds07}. All observations were
made with the ACIS-S array, with the SNR being projected
on the ACIS-S3 chip. 

Since proper motion measurements are more reliable
for longer time spans I limited the analysis to the 2000 and 2006 observations,
ignoring an additional observation made in 2004.
The 2006 observation was split in several pointings, which we
list in Table~\ref{table-obs}.
Weighing the Modified Julian Dates (MJD) of the 2006 
with the exposure times I obtain an average MJD for the 2006 observations
of 53912.13. I therefore find an average time span
between the 2006 and the 2000 \chandra\ observations of $\Delta t=5.985$~yr.

For the analysis I used the standard processed (``evt2'') events files obtained
from the \chandra\ data archive. All event files were
processed by the \chandra\ data center in 2007.

\subsection{Method}
The expansion measurements were made using the same method and 
updated C$^{++}$ code employed for the SNR
Cas A using  Einstein and ROSAT observations \citep{vink98a}. 
The results for Cas A were later confirmed by measurements 
based on \chandra\ observations \citep{delaney03}.
The method is similar to what has been used to measure
the previous X-ray expansion of Kepler \citep{hughes99}, and Tycho's
SNR \citep{hughes00c}. One aspect not explored by these studies
is the dependence of the expansion on the energy band, which was
not possible due to lack of energy resolution of Einstein and ROSAT.

For the current analysis I extracted images in
several energy bands using custom-made software, which, as described below, 
allows one
to center the image on a given sky coordinate and correct for bore sight
errors. I chose energy bands based on the spectroscopic features
of the X-ray spectrum
(see Fig.~\ref{fig-spectrum}): 
0.5-0.7 keV (covering the O VII/O VIII line emission),
0.7-1.0 keV (Fe XVII-Fe XXI L-shell emission), 
1.0-1.5 keV (Fe XXII-Fe XXIV L-shell emission, perhaps blended
with Ne IX/X and Mg XI/XII emission), 1.7-1.9 keV (Si XIII K-shell emission),
2.0-4.0 keV (covering K-shell emission from Si, S, Ar, and Ca), and
4.0-6.0 keV, which is dominated by continuum emission (a combination
of thermal bremsstrahlung and synchrotron radiation).

The expansion factor, $f$ was measured by regridding the images from
the 2006 observations,
which are statistically superior, using a simple expansion law:
\begin{eqnarray}
x_2 = {\rm round}\{\  x_0 + f (x_1 - x_0)\  + \Delta x\} \\ \nonumber
y_2 = {\rm round}\{\  y_0 + f (y_1 - y_0)\ + \Delta y\}. \label{eq-expansion}
\end{eqnarray}
with $(x_1, y_1)$  being the original pixel coordinates,
 $(x_2, y_2)$ the new pixel coordinates (rounded to the
nearest integers), and $(\Delta x,\Delta y)$,
free parameters which are fitted to correct for pointing
errors between the observations. The absolute roll angle accuracy
is very high $\sim 25$\arcsec, 
and therefore role angle errors do not contribute
to registration errors.\footnote{See 
{http://cxc.harvard.edu/cal/ASPECT/roll\_accuracy.html}, 
and  Tom Aldcroft (private communication).}
Note that pointing
errors are indistinguishable from errors in the expansion center.
For the expansion center $(x_0,y_0)$  I adopted an 
estimate of the geometrical center of the remnant based on the X-ray image:
$\alpha_{J2000} =  17^{\rm h}30^{\rm m}41.25^{\rm s}$ and 
$\delta_{J2000} = -21\degr 29\arcmin 32.95\arcsec $.

For obtaining the best fit expansion rates, maximum likelihood statistic
for Poisson distributions was used \citep{cash79}, which minimizes:
\begin{equation}
C = -2 \ln P = -2 \Sigma_{i,j} (n_{i,j} \ln m_{i,j} - m_{i,j} - \ln n_{i,j}!),\label{eq-cash}
\end{equation}
with $n_{i,j}$ being the counts in pixel $(i,j)$ of the ObsID 116 (2000) image,
and $m_{i,j}$ being the predicted counts based on the 2006 image,
which has been regridded (Eq.~\ref{eq-expansion})
and renormalized such that $\Sigma_{i,j} n_{i,j} = \Sigma_{i,j} m_{i,j}$.
The statistical fluctuations
are dominated by the observation in June 2000, with its 49 ks exposure.
I can therefore treat the combined 2006 observations with an exposure
of 750~ks as a model. Note that the last term in Eq.~\ref{eq-cash} can be
ignored for fitting purposes, because it depends only on $n_{i,j}$ which does not
change from one set of fit parameters to the other, as long as the numbers
of bins over which is summed remains unchanged.
The error in the fit parameters can be estimated using the fact that
$\Delta C = C - C_{min}$ is similar to $\Delta \chi^2$ \citep{cash79}.
The fitting procedure itself is done by scanning the relevant parameter
range, iteratively switching between the various parameters, and with each
iteration decreasing the step size.

The code has the option to fit only certain regions of the image,
using a combination of image masks and region files. For the overall fits
a mask was used, based on the broad band image, blocking out all
pixels falling below a certain threshold level. I fine tuned this mask,
such that the SNR image, smoothed with $\sigma =2$ pixels plus a border of 2 additional
pixels, falls within the mask.

\begin{figure}
\centerline{
\includegraphics[angle=-90,width=\columnwidth]{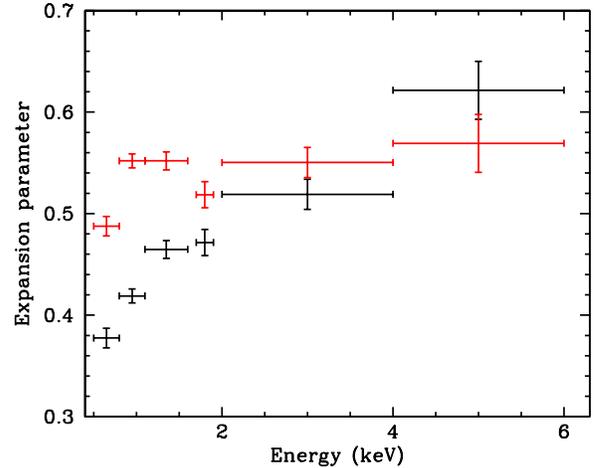}}
\caption{
The expansion parameter as function of energy. The overall expansion
rates, based on fitting the expansion for the whole remnant, are indicated
in black. The errors correspond to 90\% confidence ranges.
Due to the brightness of the northwestern region, the expansion rates are
biased toward the expansion in the  Northwest.
The expansion rate averaged over all 6 sectors are displayed in red,
with errors corresponding to the standard deviation.
\label{fig-exp_mean}
}
\end{figure}

A problem with Eq.~\ref{eq-cash} is that one can only sum over model
pixels $m_{i,j}$ that are not zero. Regridding the model image makes that
in low emissivity regions a model pixel may accidentally be set to
zero for one set of parameters, in which case it is ignored,
whereas for other values of the fit parameters it is non-zero.
Hence the number of pixels over which the statistic is derived is not
constant.
This problem does not occur for Kepler's SNR for
 energy bands with sufficient statistics, such as those covering the bright 
Fe-L emission, but it is important for the continuum image.
In order to overcome this problem the model image (based on
the 2006 observations) was smoothed with a Gaussian with $\sigma = 1$~pixel, ensuring
that the number of pixels over which the statistic is derived remained constant.
I checked for the images with the best statistics, whether smoothing 
had any effect on the expansion measurements, but within the statistical
error the small scale smoothing did not affect the measured expansion rates.
For that reason I adopted Gaussian smoothing to all 2006 images, in order to
have one consistent way of measuring for all energy bands.

The \chandra\ ACIS chips have a pixel resolution of 0.492\arcsec, slightly
undersampling the telescope point spread function.
Because of bore sight effects absolute coordinates are accurate 
up to $\sim 0.4\arcsec$. This means that by adding all the 2006 observations
one may introduce a slight blurring by approximately 1 pixel.
In order to start from the best possible images I used the same
code as for  expansion measurements reported below, but fixing $f=1$,
and using a broad band (0.3-7 keV) image. All fits were made
with respect to the ObsID 6714 image. Having fitted $(\Delta x,\Delta y)$
for each individual observation, the final extraction of the images
was made with corrections for the individual bore sights, after which
images in the same energy band were added together.
The average bore sight was 0.4 pixels in both coordinates, with rms errors
of 0.2 pixels.

Throughout this paper I use three different ways of characterizing 
the
expansion of Kepler's SNR: 
1) the expansion rate defined as $R=(1-f)/\Delta t$, with
$\Delta t = 5.985$ yr; 
2) the expansion time $\tau_{exp} = 1/R$, which is perhaps
the most intuitive number, 
as it gives the age of the remnant in case one assumes
free expansion; 
3) the expansion parameter $\beta = \tau_{\rm SN1604}/\tau_{exp}$,
the ratio of the true age of Kepler's SNR over the expansion time.

From an hydrodynamical point of view $\beta$ is the most important parameter.
In general the shock radius, $r_s$, of SNRs in distinct different
phases evolves with time $\tau$ as \citep{truelove99}:
\begin{equation}
r_s = K \tau^\beta,
\end{equation}
with $K$ a constant. This gives for the shock velocity:
\begin{equation}
v_s = \beta \frac{r_s}{\tau},
\end{equation}
this shows that
\begin{equation}
\beta = v_s \frac{\tau}{r_s} = \frac{\dot \theta d}{\theta d} \tau = \frac{(1-f)}{\Delta t} \tau = R\tau
\end{equation}
with $d$ the distance, and $\theta$ the angular radius.

\begin{figure}[b]
\centerline{
\includegraphics[angle=-90,trim=0 0 0 100,width=\columnwidth]{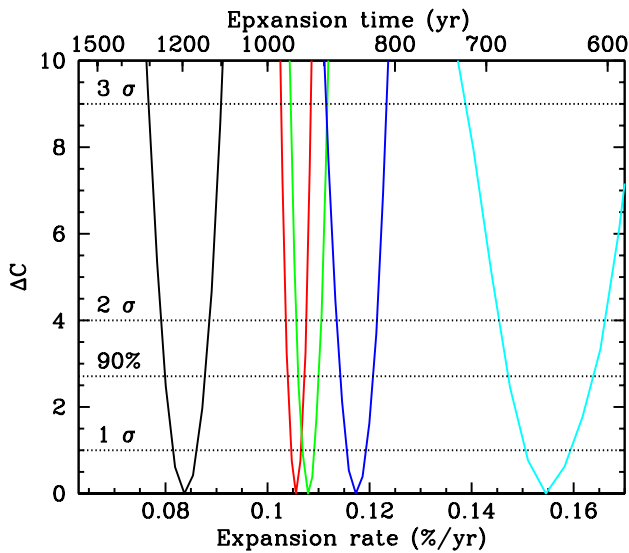}}
\caption{
Confidence ranges for the expansion rates 
for the remnant as a whole,
based on the log likelihood ratio, $\Delta C$, with respect to the best fit 
\citep{cash79}.
From left to right the curves correspond to the energy bands
0.5-0.7~keV (O VII/O VIII line emission), 
0.7-1.0~keV (Fe\-L line emission), 1.0-1.5 keV (Fe-L line emission),
1.7-1.9 keV (Si XIII line emission), and 4.0 - 6.0 keV (continuum emission).
\label{fig-deltalik}}

\end{figure}
\begin{table}
\caption{Expansion rates \label{tab-rates}}
\begin{tabular}{lllll}\noalign{\smallskip}\hline\noalign{\smallskip}

&	$R$   &		$\Delta x$&	$\Delta y$ &	$R$ ($\Delta x/\Delta y$ fixed)	\\
&	(\%/yr)&	pixels &	pixels	 & (\%/yr)	\\
\noalign{\smallskip}\hline\noalign{\smallskip}
0.5-0.7 keV &	$0.084 \pm 0.004$ & -0.58 & -0.38 & $0.084 \pm 0.004$\\
0.7-1.0 keV &	$0.108 \pm 0.002$ & -0.58 & -0.38 & $0.106 \pm 0.002$\\
1.0-1.5 keV &	$0.115 \pm 0.002$ & -0.67 & -0.45 & $0.108 \pm 0.002$\\
1.7-2.0 keV &	$0.119 \pm 0.003$ & -0.57 & -0.38 & $0.117 \pm 0.003$\\
2.0-4.0 keV &	$0.131 \pm 0.004$ & -0.74 & -0.47 & $0.129 \pm 0.004$\\
4.0-6.0 keV &	$0.158 \pm 0.006$ & -0.70 & -0.61 & $0.155 \pm 0.007$\\ 
\noalign{\smallskip}\hline\noalign{\smallskip}
Mean ($\pm$ rms) & $0.119 \pm 0.025$ & -0.64 & -0.44&$0.116\pm 0.024$ \\
$\beta$	     &  $0.48 \pm 0.10$      &     &	&	$0.47\pm 0.10$ \\
$\tau_{exp}$ &	 $840 \pm 181$	     &	   &	     &  $859 \pm 185$\\
\noalign{\smallskip}
\hline

\end{tabular}
\end{table}

\section{Results}

\subsection{The average expansion}
\label{subsec-overall}
The simplest approach to measuring the expansion of Kepler
is to fit for each energy band the expansion factor
and bore sight/expansion center offsets $(\Delta x,\Delta y)$.
Of course both the expansion center and bore sight errors
should not depend on the energy band. So a second
iteration involves fixating $(\Delta x,\Delta y)$, and then
measuring the expansion of the remnant.

However, still this is not ideal; Kepler's SNR is bright
in the northwest, so expansion measurements of the whole remnant
are skewed toward the northwest. Moreover, since a next step
involves measuring the expansion in different regions, errors
in $(\Delta x,\Delta y)$ results in errors in the expansion rate,
with opposite signs for opposite sides of the SNR.

I therefore adapted a different method: I first divided
the SNR in 4 sectors of each 90\degr. Instead of fitting
the four sectors individually, I paired them in North-South
and East-West pairs. For the North-South pair I fixated
$\Delta y$ and for the East-West pair $\Delta x$. This ensures
that there is no interference between proper motion and
bore sight corrections,
since  $(\Delta x,\Delta y)$ are measured perpendicular to
the proper motions. This procedure was repeated twice in an
iterative way for the four energy bands with sufficient
counts (the oxygen, 2 Fe-L bands, and Si band). 
The best fit values are
$(\Delta x,\Delta y)=(-0.48\pm0.01,-0.35 \pm 0.04)$, 
which were then used for
all subsequent expansion measurements.

\begin{figure*}
\centerline{
\includegraphics[width=0.97\textwidth]{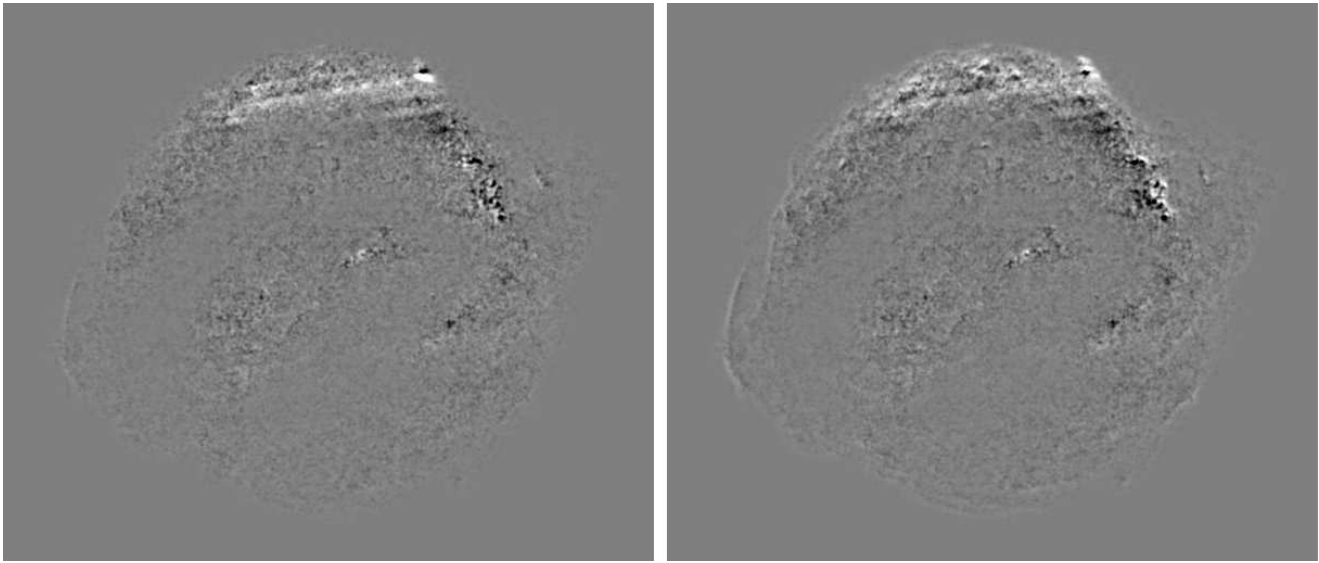}
}
\caption{
The difference between the 2000 and 2006 1.0-1.5 keV image.
A small gaussian 
smoothing has been applied to the images ($\sigma = 1$ pixel).
The image on the right 
has only been corrected for the bore sight error.
The 2006 image on the left has also been corrected for the average
expansion. 
Note that the left image still shows some expansion of the
filament in the east, and overcorrects some of the expansion in
the west (see section~\ref{subsec-azimuth}).
Also note the image artifacts in the north, which are more
visible in the left hand image, because there the expansion has been
taken out, leaving the artifacts as one of the dominant
sources of differences between the two images.
\label{fig-diff}
}
\end{figure*}

Table~\ref{tab-rates_all} lists the expansion
rates for the individual energy bands both with
$(\Delta x,\Delta y)$ as free parameters, and fixed
to the aforementioned best fit values.
Fig.~\ref{fig-exp_mean} shows the expansion parameter $\beta$ for
the individual energy bands.
Comparing the expansion rate measurements for fixed
and fitted $\Delta x/\Delta y$ shows that the measured
expansion rates are, within the statistical error, identical.
The likelihood ratios ($\Delta C$)
for the individual fits show smooth curves as a function
of expansion rate, indicating
that the minimum value for the statistic is well defined,
with no sub-minima (Fig.~\ref{fig-deltalik}).

There is a clear tendency for the expansion parameter to
increase with the photon energy , ranging from $\beta =0.34\pm 0.02$ for
the oxygen band, to $\beta = 0.62\pm 0.03$ for the continuum band.
Note that these values are significantly lower than the
expansion parameters reported by \citet{hughes99}.

Fig.~\ref{fig-diff} shows the effect of taking the average expansion
into account for the 1.0-1.5 keV band,
by showing the difference between the 2000 and 2006 images.

The difference image shows some image artifacts 
accross the northern shell. These are caused
by streaks in the June 2000 images.
The streaks are not immediately obvious in the images themselves,
but the difference image emphasizes them, in particular when
the expansion has been corrected for.
Similar artifacts showed up in measuring the expansion of Cas A with
\chandra\ \citep{delaney03}.
From the residuals after expansion correction,
I estimate that the brightness errors accross the streaks are typically
$\sim 10$\%, except near the bright knot, where the brightness error
peaks to $\sim 30$\%. Here is not quite clear whether the large deviations
are caused by the streaks or by the knots itself. I come back to the streaks
below.

\begin{figure}[b]
\centerline{\includegraphics[width=\columnwidth]{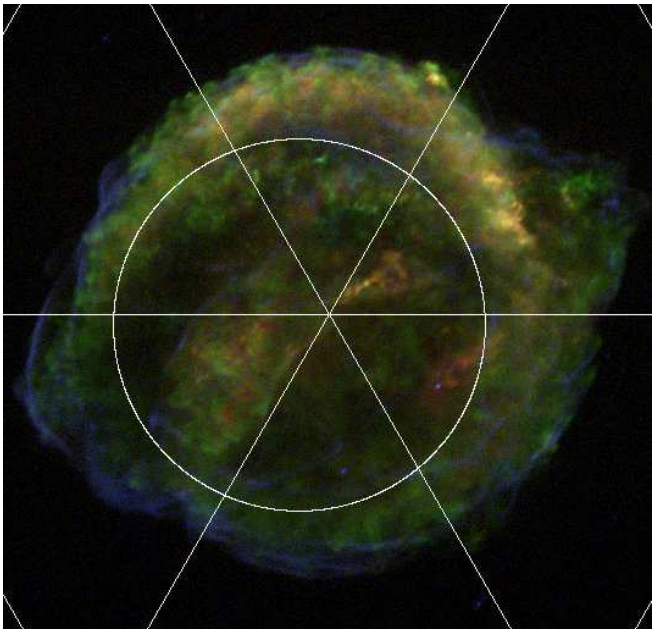}}
\caption{
Color image combining the 1.0-1.5 keV (red), 1.7-2.0 keV (green)
and 4-6 keV (blue) with overlayed the sectors used for measuring
the expansion as a function of azimuth. The circle indicates the region
not considered for the expansion measurements.
\label{fig-sectors}
}
\end{figure}

\begin{table*}
\begin{center}

\caption{Expansion rates per sector \label{tab-rates_all}}
{\scriptsize
\begin{tabular}{rlllllllll}\noalign{\smallskip}\hline\noalign{\smallskip}
&	&0.5-0.7 keV &0.7-1.0 keV & 1.0-1.5 keV &1.7-1.9 keV&2.0-4.0 keV&4.0 - 6.0 keV\\
\multicolumn{2}{l}{Sector}	  
          &     $R$ (\%/yr)    & $R$ (\%/yr)	    & $R$ (\%/yr)	  & $R$ (\%/yr)       & $R$ (\%/yr)        & $R$ (\%/yr)	\\
\noalign{\smallskip}\hline\noalign{\smallskip}

  0 &(N) &$0.074 \pm 	0.007$&$0.098 \pm 0.002$&$0.096	 \pm 0.003$&$0.105 \pm 0.005$&$0.108 \pm 0.006$&$0.080 \pm 0.015$\\
 60 &(NE)&$0.094 \pm 	0.011$&$0.133 \pm 0.004$&$0.141	 \pm 0.005$&$0.134 \pm 0.007$&$0.140 \pm 0.010$&$0.171 \pm 0.021$\\
120 &(SE)&$0.168 \pm 	0.022$&$0.164 \pm 0.010$&$0.174	 \pm 0.006$&$0.160 \pm 0.011$&$0.173 \pm 0.008$&$0.178 \pm 0.014$\\
180 &(S) &$0.155 \pm 	0.028$&$0.148 \pm 0.012$&$0.164	 \pm 0.013$&$0.143 \pm 0.014$&$0.154 \pm 0.014$&$0.109 \pm 0.016$\\
240 &(SW)&$0.116 \pm 	0.018$&$0.160 \pm 0.008$&$0.143	 \pm 0.009$&$0.139 \pm 0.011$&$0.143 \pm 0.013$&$0.173 \pm 0.021$\\
300 &(NW)&$0.068 \pm 	0.006$&$0.077 \pm 0.003$&$0.081	 \pm 0.004$&$0.093 \pm 0.007$&$0.103 \pm 0.009$&$0.138 \pm 0.021$\\
\noalign{\smallskip}
\multicolumn{2}{l}{mean ($\pm$ rms)} & 
            $0.113 \pm 0.042$   & $0.130 \pm 0.035$ & $0.133 \pm 0.037$ & $0.129 \pm 0.025$ & $0.137 \pm 0.027$ & $0.142 \pm 0.04$\\

\noalign{\smallskip}
\hline
\end{tabular}
}
\end{center}
\end{table*}

\subsection{The expansion as a function of azimuth}
\label{subsec-azimuth}
Given the strong asymmetry of Kepler's SNR it is quite
natural to expect that the expansion will also be asymmetric,
as indeed was found by \citet{dickel88} in the Radio.

I investigated this by dividing the remnant in six sectors, each spanning
60\degr in azimuthal angle (Fig.~\ref{fig-sectors}). 
In order to measure only proper motions
around the shell of the remnant, the central region was ignored,
which features a bar-like structure of unknown origin (Fig.~\ref{fig-sectors}).
For the measurements of the expansion per sector, 
the same expansion center and bore sight corrections were
applied for each sector, as for the mean expansion measurements reported
in section~\ref{subsec-overall}.

\begin{figure*}
\centerline{
\includegraphics[angle=-90,width=0.7\textwidth]{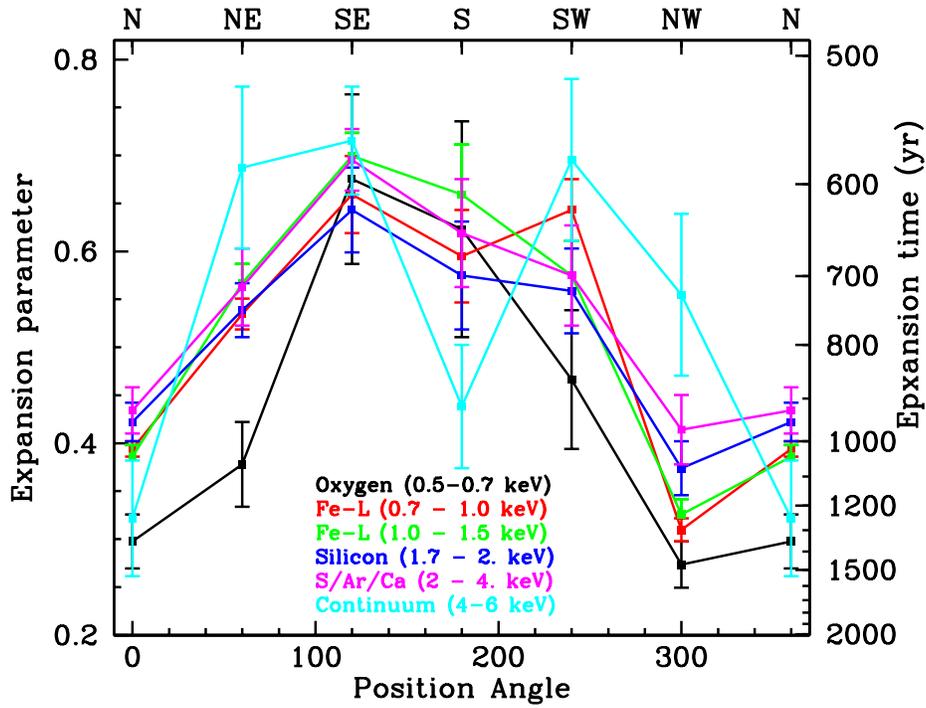}}
\caption{
The expansion parameter of Kepler's SNR
as a function of azimuthal angle. The angle is measured from the north
in a counter clockwise direction. 
The different colors indicate
expansions measured in different energy bands, using the same color
coding as in Fig.~\ref{fig-deltalik}.
The data points have been cyclic, so 
the data points at 360\degr\, repeat those at 0\degr.
The vertical axis on the right indicates the corresponding expansion time.
\label{fig-exp_sect}}

\end{figure*}

In total 36 expansion rates were measured, covering six sectors
and six X-ray bands. The expansion rates as a function
of azimuthal angle are listed in Table~\ref{tab-rates_all},
whereas the expansion parameters and expansion times are displayed in 
Fig.~\ref{fig-exp_sect}.
The expansion rates averaged over all sectors show less variation as a function
of energy than the mean expansion rates based on the whole remnant. 
An important
difference is that the averaged rates give equal weight to all sectors of the
remnant, whereas the mean expansion rate is biased toward the brighter 
northwest of the remnant (Fig.~\ref{fig-exp_mean}).

Table~\ref{tab-rates_all}, and Fig.~\ref{fig-exp_sect} show that the expansion
in the north-northwestern sectors is considerably slower than in the other
parts of the remnant: 
for the northwestern sector the expansion parameter
ranges between $\beta=0.3$ and $\beta =0.4$,
corresponding to $\tau_{exp} = 1000-1500$~yr.
For the southern and eastern sectors this is $\beta = 0.55-0.68$,  
corresponding to
$\tau_{exp}= 590-730$~yr.
Fig.~\ref{fig-exp_sect} also suggests that the variation in the average
expansion as a function of energy (Fig.~\ref{fig-exp_sect}) 
can be mostly attributed to the northwestern region.

\begin{figure*}
\centerline{
\psfig{figure=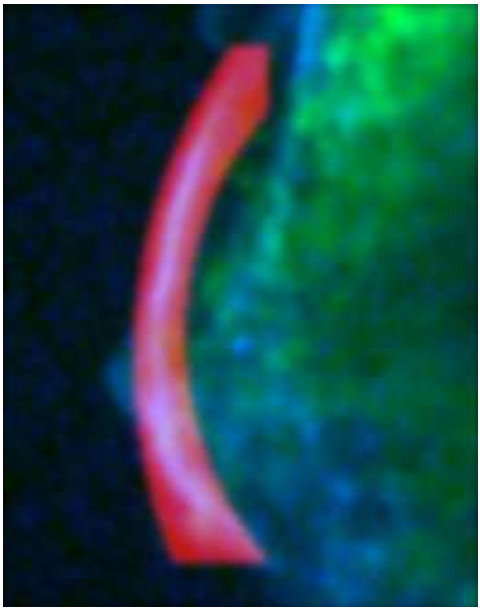,width=0.35\textwidth}
\psfig{figure=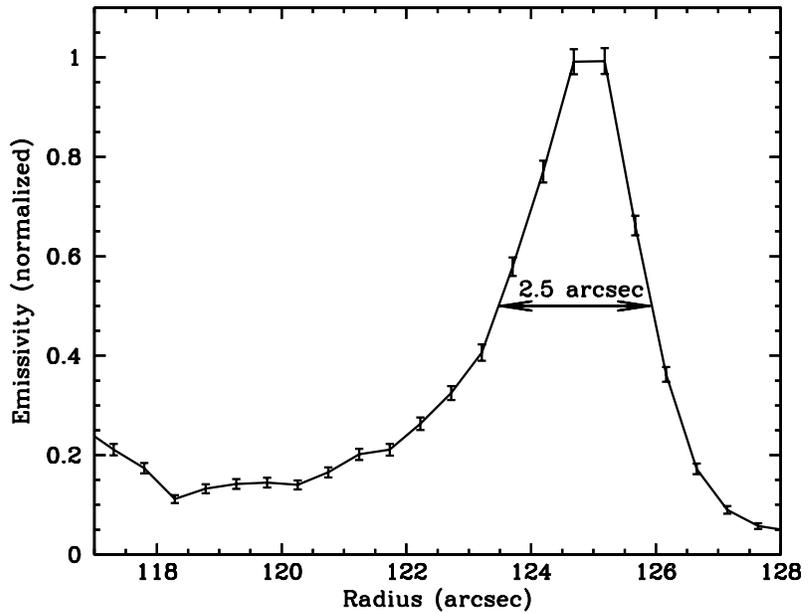,width=0.65\textwidth,angle=-90}
}
\caption{
Left:
Detail of the eastern part of Kepler's SNR, showing in
red the mask used for determining the proper motion of the
filament, in green the broad band image, and in blue
a smoothed version of 4-6 keV image.
Right: Emissivity profile of the northern region of the filament,
based on the 2006 \chandra\ observations in the 4-6 keV band.
\label{fig-filament}
}
\end{figure*}

As shown in Fig.~\ref{fig-diff} there are streaks in the June 2000 images,
which happen to be confined to the northern sector. 
The expansion measurements in this sector
gives an expansion parameter similar to that of the neigboring northwestern
sector. Given the presence of artefacts in this region one
should treat this expansion measurement with more caution than those
of the other 5 sectors. In order to have a quantitative
estimate of the streaks on the expansion measurements of the
northern region, I also measured the expansion rate after
blocking out most of the pixels affected by the streaks,
thus removing about half of the northern sector. 
Using this smaller region resulted in a higher expansion
rate, corresponding to $\beta = 0.5$, except
for the 4-6 keV band, which remained at $\beta = 0.3$. 
This difference could be either attributed to the removal of the streaks, 
i.e. the value listed in Table~\ref{tab-rates} for the northwestern region 
could be affected by systematic errors of order 30\%, 
or it could be that the regions blocked out have a slower expansion rates. 
The latter option
is quite well possible, since the streaks affect mostly
the western part of the northern sector, i.e. the region closests
to the more slowly expanding northwestern sector. In that case the different
expansion are due to a gradient accross this sector, which is 
quite plausible. Given the uncertainty, I will not explicitely discuss the
northern region, but instead concentrate on the contrast between the
slow expansion in the northwest and the rest of the remnant.

\subsection{The X-ray synchrotron filament in the Southeast}
As mentioned in the introduction, the shock velocity near 
X-ray synchrotron filaments is of considerable interest for both 
understanding magnetic field
amplification \citep{bell04,voelk05,vink06b} and the magnetic field
turbulence.

Like Cas A and Tycho, Kepler's SNR shows continuum emission
around the whole perifery of the remnant, some
of which is probably X-ray synchrotron emission \citep[e.g.][]{cassam04}. 
However, in the northwest
the continuum emission is more diffuse and associated with
regions of the most intense line radiation. It is therefore
quite likely that most, if not all, of the continuum emission
from this region is thermal
bremsstrahlung. The most unambiguous X-ray synchrotron emitting
filament is the arc-like filament in the east. As shown
by \citet{cassam04} 
and \citet{reynolds07}, 
the spectrum of this filament is completely dominated
by continuum emission. The width of this filament 
was used by \citet{voelk05} to
infer a magnetic field of 250~$\mu$G \citep[see also][]{bamba05}.
The 2006 observations also show that the filament is narrow,
2.5\arcsec. However, note that the estimate of the magnetic
field by \citet{voelk05} assumes spherical symmetry in order to
estimate the true, deprojected, filament width. Taking
the 2.5\arcsec at face value results in a magnetic field 
of 85~$\mu$G using the formula in \citet{vink06d}.
This should be taken as a conservative lower limit.

Since it is important to know under what conditions
X-ray synchrotron emission occurs,
I measured the expansion of the eastern filament separately. 
The spectrum from
this filament shows hardly any line emission \citep{reynolds07}, so
I measured the expansion using a 0.3-7 keV broad band image
in order to improve the statistics of the measurements.
The expansion rate that is found is $R= (0.176\pm 0.007)$\%,
corresponding to a relatively large expansion parameter of 
$\beta=0.71\pm 0.03$.
The filament is located at an angular radius of 2.1\arcmin\
from the center.
Since X-ray synchrotron filaments trace the shock front, one can therefore
translate the expansion parameter into a shock velocity
of $v_s = (4200 \pm 170) d_4$~\kms,
with $d_4$ the distance in units of 4~kpc. This is twice as
fast as the shock velocities inferred from optical spectral and 
proper motions studies in the northwestern region
\citep{blair91,sankrit05}, but consistent with
the value adopted by \citet{voelk05}.

Since the shape of the filament has a radius of curvature
smaller than the radius of the remnant, one may wonder
in what directions the filament is actually expanding:
Is the filament caused by a blow out, in which case
one expects the expansion center to be closer to the
approximate curvature center of the filament, or is
the expansion center close to the geometrical center
of the whole remnant?

In order to get some handle on this, I also fitted the expansion, but
leaving the center of expansion as free parameters.
In that case, the best fit center of expansion was 
more toward the west ($\alpha_{J2000} =  17^{\rm h}30^{\rm m}40.77^{\rm s}$,
$\delta_{J2000} = -21\degr 29\arcmin 29.49\arcsec $) than the adopted center, i.e. opposite
of the non-thermal filament. This, and in addition blinking of
the 2000 and 2006 images by eye, suggests that the curved
structure is moving more or less as a coherent structure,
rather than expanding from a center close to the filament.
This is reminiscent of the kinematics of a bow shock structure.
The expansion parameter did not change substantially, when the
expansion center was treated as a free parameter: 
$\beta = 0.67\pm 0.04$.

\section{Discussion}

I have measured the expansion of Kepler's SNR using archival \chandra\
data from observations performed in 2000 and 2006. These new
X-ray expansion measurements largely agree with expansion measurements
based on radio \citep{dickel88} and optical \citep{sankrit05} measurements.
Specifically, the results confirm that the average expansion parameter is 
$\beta \approx 0.5$.
The expansion as a function
of azimuthal angle shows a clear difference in expansion rate between the 
northwestern
and other parts of the remnant with the northwestern part having an
expansion parameter $\beta \approx 0.3-0.4$, as also found
in the radio \citep{dickel88} and optical  \citep{sankrit05},
and the other parts having  $\beta \approx 0.6$, in agreement
with the radio measurements.

The expansion measurements presented here
are in disagreement with previous X-ray measurements,
based on ROSAT and Einstein data \citep{hughes99}, which suggested
$\beta \approx 0.9$. In terms of resolution this new measurement
should be better than the ROSAT-Einstein measurement, despite 
the long baseline of the latter, 17.5~yr. 
The resolution of the high resolution imagers
on board the ROSAT and Einstein satellites is about 4\arcsec,
amounting to 0.22\arcsec\,yr$^{-1}$\ for a baseline of 17.5~yr. 
The pixel resolution of the
\chandra\ ACIS instrument is 0.43\arcsec, so the resolution per unit time
for the present study is about 0.07\arcsec\, yr$^{-1}$.
It is difficult to assess what causes the discrepancy
between the new result
and the ROSAT/Einstein study, as the measurements
by \citet{hughes99} used a similar method as employed here, and
for the expansion of Cas A 
\citep[][also based on Einstein and ROSAT data]{vink98a}.

\subsection{The expansion parameter in theoretical models}
Theoretical models of SNR evolution predict directly the
expansion parameter, a dimensionless number 
\citep{chevalier82,dwarkadas98,truelove99}.
The best known example is  the evolution of the shock wave in the so called
Sedov-Taylor phase, which treats the supernova as a point explosion
in a uniform density medium. This gives $\beta = 0.4$.
\citet{chevalier82} analyzed the early evolution of a SNR, in the context
of power law ejecta density models, i.e. $\rho \propto v^{-n}$. 
This gives $\beta = (n-3)/n$ for an explosion in an uniform density medium.
For Type Ia supernovae it has been argued that $n=7$, which should
therefore result in $\beta = 0.57$\ during the early phase of the SNR. 

The expansion parameter in this case refers to both the contact discontinuity
(separating shocked circumstellar medium (CSM) 
and shocked ejecta) and the forward shock.
The plasma in the shell has a range of expansion parameters centered around 
$\beta = 0.57$, ranging from $\beta = 0.63$ near the reverse shock,
to $\beta = 0.43$ near the forward shock.

\citet{dwarkadas98} studied the hydrodynamical 
evolution  of Type Ia remnants numerically, using both power law
ejecta density profiles, and exponential ejecta density profile
($\rho \propto \exp( -v/v_e)$). They 
explicitly provide the expansion parameter
of the shock itself, which in case of the exponential density profile
ranges from $\beta =0.8$ very early in the evolution to $\beta = 0.4$,
although never reaching $\beta = 0.4$. The expansion parameter
of the shocked plasma can be obtained from their Fig.~3: At a late phase,
around $t \approx 500 E_{51}^{-0.5}n_0^{-1/3}$~yr ($n_0$ being the preshocked
number density and $E_{51}$ the explosion energy in units of $10^{51}$~erg) 
for both types of density
profiles $\beta \approx 0.5-0.6$ for the forward shock, but for the
plasma the expansion parameter is almost uniformly $\beta \approx 0.38$. 
The lower expansion parameter for the plasma is not surprising, given that 
the plasma directly behind the shock moves with $v = 3v_s/4$  in the
case of a monatomic gas with adiabatic index $\gamma=5/3$. Therefore,
in the late phase $\beta_{shock} = 4\beta_{plasma}/3$.

These studies are important for interpreting the measured expansion parameters
of Kepler's SNR. First, it is good to be aware that the expansion parameters
of the shock may be different from the plasma. In the late phase, the plasma
expansion parameter is lower than that of the shock itself, whereas
in the early phase of the evolution there is a range of values, but
around the contact discontinuity the expansion parameter is similar to that
of  the shock. Therefore, during the early evolution
the expansion parameter is expected to be close to the expansion parameter
of the forward shock.

\subsection{Inferences on the shock velocities}
The question now arises what the X-ray expansion measurements really provides:
the expansion parameter of the shock or that of the plasma behind it?
The shock velocity
is a pattern speed, and this is certainly part of what is measured.
As Fig.~\ref{fig-diff} shows, correcting for the mean expansion removes 
the strong
fringes in the difference map,
around the forward shock.
On the other hand, also the velocity
of the plasma  itself may influence the best fit expansion parameters.
The measurements themselves are skewed toward the outer part of the shell,
simply because the proper motions are larger there, and more pixels
are involved. 

The models discussed above therefore suggest that the measured
expansion parameter is a lower limit to the expansion parameter
of the forward shock, later in the evolution of
the SNR 
\citep[i.e. ~$t^\prime > 1$ in the notation of][their Eq.~6]{dwarkadas98}.
In an earlier phase, for which the \citet{chevalier82} models may
be applicable, the measured expansion parameter may be a good representation
of the expansion parameter of the forward shock.

The expansion measurement for most of the remnant, except for the northwestern
part, gives  $\beta \approx 0.6$, which is consistent with the expansion
parameter of $\beta=0.57$ for both the contact discontinuity and 
the forward shock for
the $n=7$ model of \citet{chevalier82}, the prefered
model for Type Ia supernovae.
It is also in the approximate range for the forward shock expansion parameter
for the exponential density
profile model considered by \citet{dwarkadas98}, but only
in case the remnant is still in the early phase of its evolution,
i.e. if the self-similar time variable is $t^\prime \lesssim 1$.

Note that the theoretical values for the expansion parameter, whether
for the shock or the plasma, are all lower than the expansion parameters
obtained from the previous X-ray measurements based on ROSAT and
Einstein observations \citep{hughes99}, namely $\beta = 0.93$.
The \citet{chevalier82} model for $n=7$ gives for the plasma velocities
a range of $\beta = 0.4-0.6$, whereas the exponential density profiles
gives $\beta < 0.8$ after the first few decades in the life of the SNR
\citep{dwarkadas98}.
In addition,the previous X-ray  expansion measurements
are inconsistent with the Fe-K  emission detected to come from
reverse shocked material in Kepler's SNR, as discussed by \citet{cassam04};
Fe-K emission requires a well advanced reverse shock and 
high reverse shock speed, which is incompatible with free expansion.

The value for the expansion parameter in the southern and eastern
sectors are consistent with the expansion parameter for
the $n=7$ model of  \citet{chevalier82}, which applies
to both the contact discontuity and the forward shock.
It seems, therefore, justified to translate the measured
expansion parameter into a shock velocity, using that the average
angular radius of  Kepler is $r=1.76$\arcmin. This gives
$v_s = 3010 (\beta/0.6) d_4$~\kms, with $d_4$ the distance
in units of 4~kpc.

For the northwestern region shock velocities have been measured
using the widths of  H$\alpha$ lines, which gives
a direct measurement of the post-shock proton temperature.
These suggest shock velocities in this region of 1500-2800~\kms 
\citep{fesen89,blair91}, with the most likely value centered
around 1660~\kms\ \citep{sankrit05}. The proper motions of the 
H$\alpha$\ emitting shock regions is 0.088\arcsec~yr$^{-1}$ 
\citep{sankrit05}. This corresponds to an optical expansion parameter of
$\beta = 0.33$. This value is consistent with the value
reported here for the X-ray expansion parameter for the
northwestern region, $\beta=0.3-0.4$, and with the radio expansion
parameter $\beta \approx 0.35$. 

Finally, the highest expansion reported in this paper is for the X-ray 
synchrotron filament in the east ($\beta = 0.71$).
The size of the X-ray synchrotron emitting region is determined by
the loss time of the highest energy electrons and their advection away of
the shock front \citep{vink03a,vink06b}. As a result the size of the
X-ray emitting region is expected to be fixed as a long as the magnetic
field and the shock speed are approximately constant. The displacement
of the filament therefore reflects the speed of the shock front, rather than
the movement of the plasma.
This suggests that the measured filament movement is a pattern movement,
and reflects directly the movement of the shock front,
implying $v_s= 4200 d_4$~\kms.

This relatively high shock velocity  and the 
protrusion of the filament outside the general shock radius
of Kepler's SNR make it quite likely that the filament
marks a part of the shock that is expanding into a low density
region, as previously suggested by \citet{cassam04}.
Moreover, the high velocity and the low density both help to
explain why in this region X-ray synchrotron emission dominates
over thermal emission. The high velocity is consistent with
the idea that only shocks with velocities $v_s \gtrsim 2000$~\kms
give rise to detectable X-ray synchrotron radiation 
\citep{aharonian99,vink06d,helder08}.

\begin{figure}
\centerline{\includegraphics[width=\columnwidth]{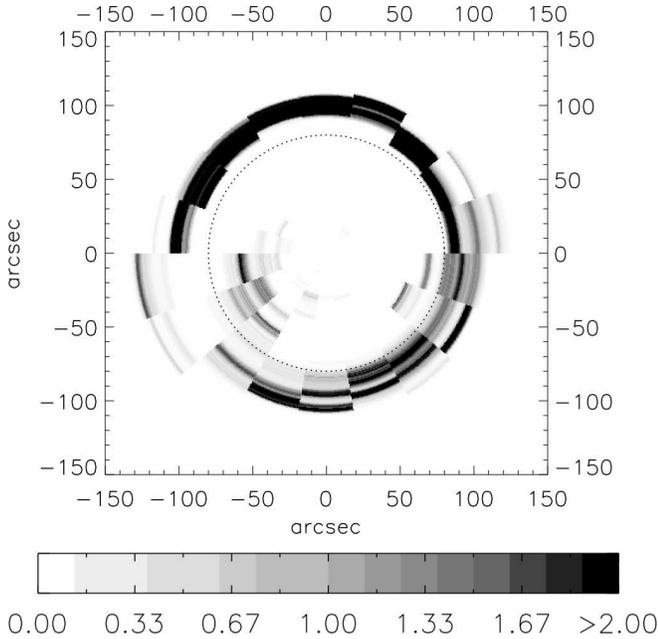}}
\caption{
The deprojected 1-1.5 keV (Fe-L) image of Kepler's supernova remnant,
used to estimate the location of the reverse shock
(indicated by the dotted line with a radius of 1.3\arcmin).
The deprojection was made in 18 independent sectors, following
the procedure described in \citet{helder08}.
The brightness scale is in percentage per bin, scaled in such a way that the 
total adds up to 100\% integrated flux in each sector. 
(Figure kindly provided by Eveline Helder.)
\label{fig_helder}
}
\end{figure}

\begin{figure*}
\centerline{
\includegraphics[angle=-90,width=0.9\textwidth]{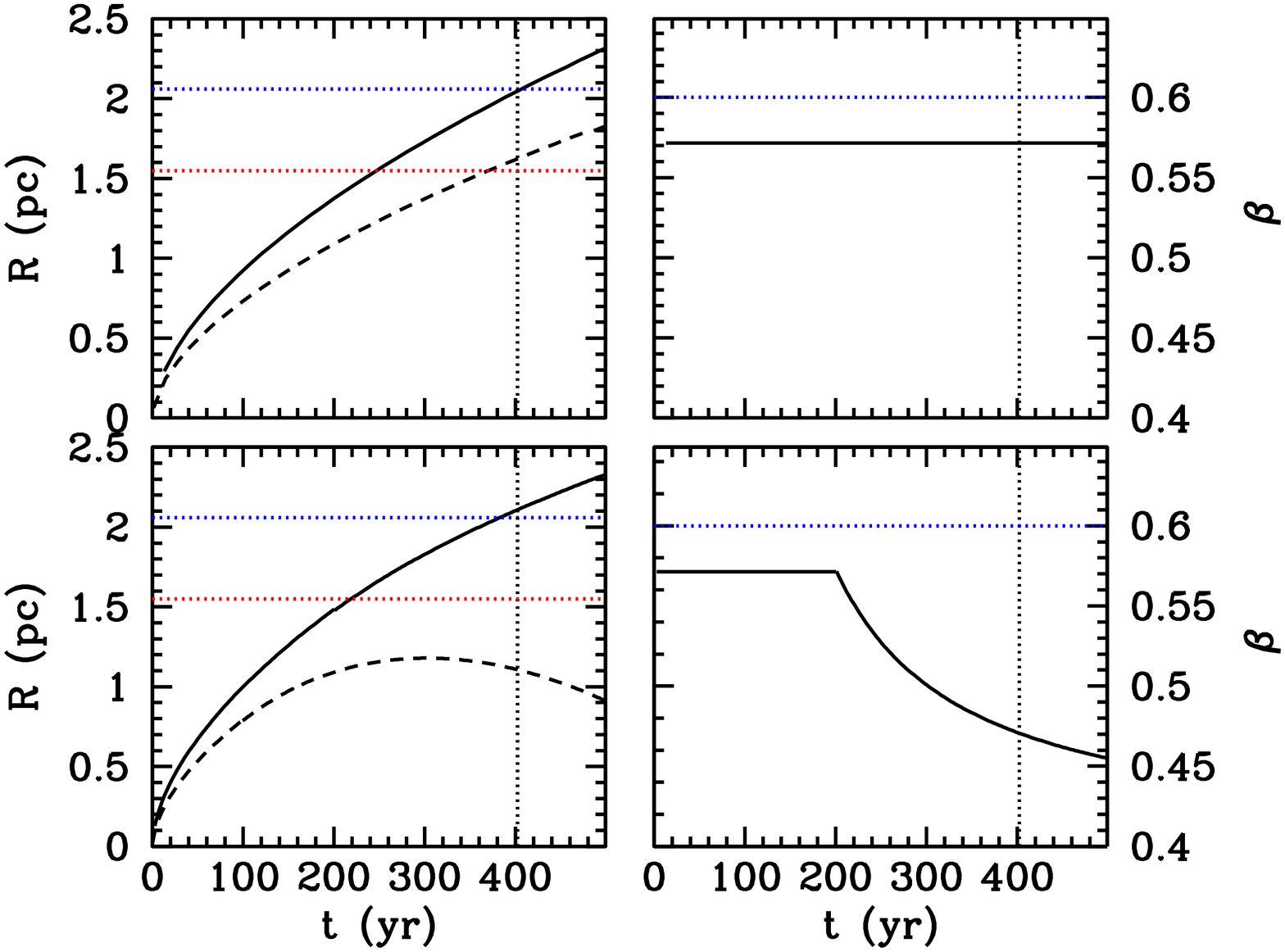}}
\caption{
The evolution of shock radii (left) and expansion parameters of the foward
shock (right) according to the $n=7$ model
of \citet{truelove99} for two different sets of kinetic energies
and circumstellar densities. The values were adapted such that the
forward shock matches 2.06~pc (indicated by the upper dotted line)
at the age of Kepler's SNR (402 yr),
valid for a distance of 4~kpc.
In the top panels $n_H = 1$~cm$^{-3}$, resulting in a kinetic energy of
$E_0 = 2\times 10^{50}$~erg.
In the bottom panels $E_0 = 10^{51}$~erg, in which case $n_H = 12$~cm$^{-3}$.
In this case both the position of the reverse shock (the lower dotted line
in the left panels) and the expansion
parameter do not match the measurements.
\label{fig-truelove}
}
\end{figure*}

\subsection{The expansion and the energetics of SN1604}
The regions of Kepler's SNR that are likely to be most revealing about the
explosion parameters of SN1604 are the southern/southwestern regions. 
In the northwestern
region it is clear that the expansion parameter is too low to fit
either the models of \citet{chevalier82} and \citet{dwarkadas98}, or
the Sedov evolution. In the east the radius of the remnant is not well 
determined, due to the protruding non-thermal filament.
For the southwestern region the shock radius is 1.77\arcmin, corresponding
to a physical radius of $r_s =2.06 d_4$~pc. As an additional constraint we
can use the radius of the reverse shock. From the 1-1.5 keV (Fe-L) image 
the typical reverse shock radius
is estimated to be $\sim 1.3$\arcmin\ (Fig.~\ref{fig_helder}),
corresponding to a reverse shock radius of $r_r \approx 1.6 d_4$~pc.
Using these values, and the measured expansion parameters for the 
southern/southwestern regions ($\beta \approx 0.6$) in conjunction with
the models of \citet{truelove99} or \citet{dwarkadas98}, we can
constrain the pre-shock density and explosion energy for Kepler's SNR.

Fig.~\ref{fig-truelove} 
shows the $n=7$ \citet{truelove99} model for two different
choices of kinetic energy and circumstellar medium density,
assuming a distance of 4~kpc. 
One choice is to assume that $n_{\rm H} = 1$~cm$^{-3}$,
the other that the energy
is the canonical explosion
energy of $E_0 = 10^{51}$~erg ($\equiv 1$ Bethe). In both cases
it has been assumed that the ejecta mass is $M_{ej} = 1.4$~\msun,
as is to be expected for a Type Ia supernova \citep[][for recent discussions
on this issue]{woosley07,mazzali07}.
It is clear that both in terms of the reverse shock position in Kepler's SNR
and in terms of measured expansion parameters a lower kinetic energy is to be
prefered.

Comparing the measured properties for the southern part of Kepler's SNR with
the numerical models of \citet{dwarkadas98} leads to a similar conclusion.
The measured expansion parameters are consistent with a low value
of the self-similar time coordinate in \citet{dwarkadas98}, 
i.e. $t^\prime \lesssim 1$, with $\beta \approx 0.6$, corresponding
to $t^\prime \approx 0.6$. However, a conservative upper limit is 
$t^\prime < 2$,
corresponding to $\beta = 0.5$ for the shock itself.
Translating the dimensionless time, and the associated normalized shock radii
($0.85 \lesssim r^\prime \lesssim 1.6$) into a physical ages and shock radius
using the conversion equations in \citet[][Eq.4-6]{dwarkadas98}, 
one finds
that $0.7 \lesssim n_{\rm H}/({\rm cm}^{-3}) \lesssim 5$, and
$0.2 \lesssim E_0/({\rm 1 B}) \lesssim 0.5$, with a preference for
the lower values.

So for both the \citet{truelove99} and \citet{dwarkadas98} models
it appears that SN1604 had a relatively low explosion energy.
However, this depends also on the adopted distance of 4~kpc 
\citep{sankrit05}. 
For 5~kpc, 
close to the nominal distance estimate of \citet{reynoso99}, the allowed
range of densities and energies is 
$0.4 \lesssim n_{\rm H}/({\rm cm}^{-3}) \lesssim 2.5$, and
$0.3 \lesssim E_0/({\rm 1 B}) \lesssim 0.9$, which is still
rather low compared to the uniform kinetic energies of $1.2\pm 0.2$~B
inferred for observed Type Ia supernovae \citep{woosley07}.
Moreover, the upper limit requires a relatively high density for a SNR
located $470 d_4$~pc above the Galactic plane.
Only for an adopted distance of 6~kpc is the angular radius 
of the forward and reverse shock
of the SNR consistent with a kinetic energy
of $10^{51}$~erg. The associated ISM density is then 
$n_{\rm H} = 0.5 ({\rm cm}^{-3})$.
A distance considerably further than 4~kpc seems therefore preferable,
provided that
SN1604 was indeed a Type Ia supernova, and
considering the evidence that most Type Ia supernovae have energies
in excess of  $10^{51}$~erg \citep{woosley07}. 

A similar conclusion
was recently obtained based on the non-detection of TeV emission from
Kepler with the H.E.S.S. telescope \citep{aharonian08}. 
Note that the conclusion of \citet{aharonian08}
depends on assumptions concerning the TeV luminosity of the remnant,
which depends on the explosion energy and on the fraction of the energy
that goes into accelerating cosmic rays.

\subsection{Estimates of the swept up mass in the Northwest}
The northwestern region of Kepler's SNR has an expansion
parameter even lower than the expansion parameter expected for
the Sedov-Taylor phase. 
This is probably caused by
a non-uniform density profile. The northwestern part of the remnant shows
in the optical nitrogen-rich material, suggesting a shell ejected by
the progenitor system. The fact that the expansion rate in the northwestern
region seems to
be a function of photon energy may be accounted for if the
inner layer of shocked ejecta is hotter, and if in this layer
the expansion parameter is larger. The slowest expansion is measured
for the oxygen band, and it has been suggested that most oxygen
emission is from shocked circumstellar medium, rather than from
the ejecta \citep{cassam04}.
The trend with energy in this region could be an effect of a slow
response of the inner layers to density enhancements closer to the shock.
This reinforces the idea of a dense shell that has recently been encountered
by the blastwave.

As noted by \citet{reynolds07} the presence of nitrogen rich material from the
progenitor system is difficult, but not impossible, to reconcile with SN1604
being a Type Ia supernova. A possibility is, for example,
that the supernova belonged to the short-lived Type Ia channel 
\citep{mannucci06}, in which the white dwarf progenitor and 
its companion star, were relatively massive stars.
However, for the remnant of SN1604 the complication remains
that the progenitor, or its companion, must have deposited
a substantial amount of mass at a large distance.

The amount of material must have been substantial in order
for the shock to have decelerated so much that the expansion
parameter is even lower than expected for the Sedov-Taylor phase.
If we assume the material encountered in the northwest was a shell,
covering a fraction of about $f=0.25$ one can estimate the mass in
the shell by requiring that the mass in the shell must be more
than the swept up interstellar medium in other parts of the remnant,
which have $n_{\rm H} \approx 1$~cm$^{-3}$. This corresponds to a 
mass of $M_{swept} = 1 (f/0.25) n_{\rm H} d_4^3$~\msun.
It therefore seems reasonable to assume that the mass encountered
in the northwest is also about 1~\msun.

This material must have been 
lost from either the progenitor of the supernova,
or from its companion star. This 
suggests a strongly non-conservative
binary evolution scenario: Too much mass in the shell
means less mass available for accretion onto the white dwarf,
complicating its evolution toward a Type Ia supernova.
Another problem may be how to eject this material to
a distance of of $\sim 2$~pc from the  progenitor. Perhaps,
the shell is caused by nova explosions on the progenitor?

The remnant of SN1604 remains, therefore, a puzzling, but intriguing object.
Due its unusual properties it may in the future reveal new aspects
of Type Ia supernovae. Future studies of the SNR may provide some answers,
but a high priority would be to identify light echos of SN1604,
and obtain their optical spectra.
This has been done very recently for the SNRs Cas A \citep{krause08},
which appears to have been a Type IIb supernova,   and
SNR 0509-67.5, an energetic Type Ia supernova in the Large Magellanic Cloud
\citep{rest08a}.

\section{Summary}
The expansion of Kepler's SNR (SN1604) was measured using
archival \chandra\ data. The expansion
in all parts of the remnant is inconsistent with
free expansion with the expansion measurement of $\beta \approx 0.9$,
that was previously reported \citep{hughes99}.

The X-ray measurements reported here,
and  previous radio and optical expansion
measurements, show that the remnant expands more slowly in the bright
northwestern part, $\beta \approx 0.3 -0.4$,
than in the rest of the supernova remnant, where
$\beta \approx 0.6$.
The fastest expansion is found for the X-ray synchrotron filament
in the eastern part of the remnant, $\beta \approx 0.7$.

The remnant seems not yet to  have entered
the Sedov-Taylor phase of its evolution:
Apart from the northwestern region
the expansion parameters are consistent with the early expansion phase
as detailed in the models by \citet{chevalier82,truelove99,dwarkadas98}.
For the northwestern part of the remnant a different scenario
is needed, since it's expansion parameter is even smaller
than for a Sedov-Taylor evolution model.

The kinematics of Kepler's SNR, in particular the more undisturbed
southwestern region, is only consistent with current
hydrodynamical models of Type Ia SNRs,
if its distance is considerably larger
than the 4~kpc obtained by \citet{sankrit05}. Recently,
\citet{aharonian08} suggested that $d > 6$~kpc, based on the non-detection
of Kepler by the H.E.S.S. TeV $\gamma$-ray telescope.
Both in the present paper and in \citet{aharonian08}, the conclusion
regarding a large distance is based on the assumption that Kepler was indeed
a Type Ia SNR with an explosion energy $\gtrsim 10^{51}$~erg.

The X-ray synchrotron filament seems to move with a shock speed
of $4200 d_4$~\kms, consistent with theory \citep{aharonian99}
and other observations \citep{helder08} that indicate that 
only shocks with 
$v_s \gtrsim 2000$~\kms\ emit X-ray synchrotron radiation.

\acknowledgements
This work is supported by a Vidi grant
from the Netherlands Organisation
for Scientific Research (NWO).
I would like to thank Eveline Helder,
Klara Schure,
Frank Verbunt,
and Daria Kosenko for discussions and
helpful comments on the manuscipt.
In addition I thank Eveline Helder for providing me
with Fig.~\ref{fig_helder}.


\begin{thebibliography}{40}
\expandafter\ifx\csname natexlab\endcsname\relax\def\natexlab#1{#1}\fi

\bibitem[{{Aharonian} {et~al.}(2008)}]{aharonian08}
{Aharonian}, F. {et~al.} 2008, A\&A accepted (ArXiv:0806.3347)

\bibitem[{{Aharonian} \& {Atoyan}(1999)}]{aharonian99}
{Aharonian}, F.~A. \& {Atoyan}, A.~M. 1999, \aap, 351, 330

\bibitem[{{Baade}(1943)}]{baade43}
{Baade}, W. 1943, \apj, 97, 119

\bibitem[{{Bamba} {et~al.}(2005)}]{bamba05}
{Bamba}, A. {et~al.} 2005, \apj, 621, 793

\bibitem[{{Bandiera}(1987)}]{bandiera87}
{Bandiera}, R. 1987, \apj, 319, 885

\bibitem[{{Bell}(2004)}]{bell04}
{Bell}, A.~R. 2004, \mnras, 353, 550

\bibitem[{{Bell} \& {Lucek}(2001)}]{bell01}
{Bell}, A.~R. \& {Lucek}, S.~G. 2001, \mnras, 321, 433

\bibitem[{{Blair}(2005)}]{blair05}
{Blair}, W.~P. 2005, in Astronomical Society of the Pacific Conference Series,
  Vol. 342, 1604-2004: Supernovae as Cosmological Lighthouses, ed.
  M.~{Turatto}, S.~{Benetti}, L.~{Zampieri}, \& W.~{Shea}, 416--+

\bibitem[{{Blair} {et~al.}(1991){Blair}, {Long}, \& {Vancura}}]{blair91}
{Blair}, W.~P., {Long}, K.~S., \& {Vancura}, O. 1991, \apj, 366, 484

\bibitem[{{Cash}(1979)}]{cash79}
{Cash}, W. 1979, \apj, 228, 939

\bibitem[{{Cassam-Chena{\" i}} {et~al.}(2004)}]{cassam04}
{Cassam-Chena{\" i}}, G. {et~al.} 2004, \aap, 414, 545

\bibitem[{{Chevalier}(1982)}]{chevalier82}
{Chevalier}, R.~A. 1982, \apj, 258, 790

\bibitem[{{Delaney} \& {Rudnick}(2003)}]{delaney03}
{Delaney}, T. \& {Rudnick}, L. 2003, \apj, 589, 818

\bibitem[{{Dickel} {et~al.}(1988){Dickel}, {Sault}, {Arendt}, {Korista}, \&
  {Matsui}}]{dickel88}
{Dickel}, J.~R., {Sault}, R., {Arendt}, R.~G., {Korista}, K.~T., \& {Matsui},
  Y. 1988, \apj, 330, 254

\bibitem[{{Dwarkadas} \& {Chevalier}(1998)}]{dwarkadas98}
{Dwarkadas}, V.~V. \& {Chevalier}, R.~A. 1998, \apj, 497, 807

\bibitem[{{Fesen} {et~al.}(1989){Fesen}, {Becker}, {Blair}, \&
  {Long}}]{fesen89}
{Fesen}, R.~A., {Becker}, R.~H., {Blair}, W.~P., \& {Long}, K.~S. 1989, \apjl,
  338, L13

\bibitem[{{Helder} \& {Vink}(2008)}]{helder08}
{Helder}, E.~A. \& {Vink}, J. 2008, ApJ accepted (ArXiv:0806.3748)

\bibitem[{{Hughes}(1999)}]{hughes99}
{Hughes}, J.~P. 1999, \apj, 527, 298

\bibitem[{{Hughes}(2000)}]{hughes00c}
---. 2000, \apjl, 545, L53

\bibitem[{{Kinugasa} \& {Tsunemi}(1999)}]{kinugasa99}
{Kinugasa}, K. \& {Tsunemi}, H. 1999, \pasj, 51, 239

\bibitem[{{Koralesky} {et~al.}(1998){Koralesky}, {Rudnick}, {Gotthelf}, \&
  {Keohane}}]{koralesky98}
{Koralesky}, B., {Rudnick}, L., {Gotthelf}, E.~V., \& {Keohane}, J.~W. 1998,
  \apjl, 505, L27+

\bibitem[{{Krause} {et~al.}(2008){Krause}, {Birkmann}, {Usuda}, {Hattori},
  {Goto}, {Rieke}, \& {Misselt}}]{krause08}
{Krause}, O., {Birkmann}, S.~M., {Usuda}, T., {Hattori}, T., {Goto}, M.,
  {Rieke}, G.~H., \& {Misselt}, K.~A. 2008, Science, 320, 1195

\bibitem[{{Malkov} \& {Drury}(2001)}]{malkov01}
{Malkov}, M.~A. \& {Drury}, L. 2001, Reports of Progress in Physics, 64, 429

\bibitem[{{Mannucci} {et~al.}(2006){Mannucci}, {Della Valle}, \&
  {Panagia}}]{mannucci06}
{Mannucci}, F., {Della Valle}, M., \& {Panagia}, N. 2006, \mnras, 370, 773

\bibitem[{{Mazzali} {et~al.}(2007){Mazzali}, {R{\"o}pke}, {Benetti}, \&
  {Hillebrandt}}]{mazzali07}
{Mazzali}, P.~A., {R{\"o}pke}, F.~K., {Benetti}, S., \& {Hillebrandt}, W. 2007,
  Science, 315, 825

\bibitem[{{Rest} {et~al.}(2008)}]{rest08a}
{Rest}, A. {et~al.} 2008, astro-ph, 8014762

\bibitem[{{Reynolds} {et~al.}(2007){Reynolds}, {Borkowski}, {Hwang}, {Hughes},
  {Badenes}, {Laming}, \& {Blondin}}]{reynolds07}
{Reynolds}, S.~P., {Borkowski}, K.~J., {Hwang}, U., {Hughes}, J.~P., {Badenes},
  C., {Laming}, J.~M., \& {Blondin}, J.~M. 2007, \apjl, 668, L135

\bibitem[{{Reynoso} \& {Goss}(1999)}]{reynoso99}
{Reynoso}, E.~M. \& {Goss}, W.~M. 1999, \aj, 118, 926

\bibitem[{{Sankrit} {et~al.}(2005){Sankrit}, {Blair}, {Delaney}, {Rudnick},
  {Harrus}, \& {Ennis}}]{sankrit05}
{Sankrit}, R., {Blair}, W.~P., {Delaney}, T., {Rudnick}, L., {Harrus}, I.~M.,
  \& {Ennis}, J.~A. 2005, Advances in Space Research, 35, 1027

\bibitem[{{Stage} {et~al.}(2006){Stage}, {Allen}, {Houck}, \&
  {Davis}}]{stage06}
{Stage}, M.~D., {Allen}, G.~E., {Houck}, J.~C., \& {Davis}, J.~E. 2006, Nature
  Physics, 2, 614

\bibitem[{{Stephenson} \& {Green}(2002)}]{stephenson02}
{Stephenson}, F.~R. \& {Green}, D.~A. 2002, {Historical supernovae and their
  remnants} (Oxford: Clarendon Press)

\bibitem[{{Truelove} \& {McKee}(1999)}]{truelove99}
{Truelove}, J.~K. \& {McKee}, C.~F. 1999, \apjs, 120, 299

\bibitem[{{V{\" o}lk} {et~al.}(2005){V{\" o}lk}, {Berezhko}, \&
  {Ksenofontov}}]{voelk05}
{V{\" o}lk}, H.~J., {Berezhko}, E.~G., \& {Ksenofontov}, L.~T. 2005, \aap, 433,
  229

\bibitem[{{Vink}(2004)}]{vink04b}
{Vink}, J. 2004, \adspr, 33, 356

\bibitem[{{Vink}(2006)}]{vink06b}
{Vink}, J. 2006, in The X-ray Universe 2005, ESA SP-604 Vol. 1, A. Wilson ed.
  (ESA, ESTEC, The Netherlands), {319}

\bibitem[{{Vink} {et~al.}(2006){Vink}, {Bleeker}, {van der Heyden}, {Bykov},
  {Bamba}, \& {Yamazaki}}]{vink06d}
{Vink}, J., {Bleeker}, J., {van der Heyden}, K., {Bykov}, A., {Bamba}, A., \&
  {Yamazaki}, R. 2006, \apjl, 648, L33

\bibitem[{{Vink} {et~al.}(1998){Vink}, {Bloemen}, {Kaastra}, \&
  {Bleeker}}]{vink98a}
{Vink}, J., {Bloemen}, H., {Kaastra}, J.~S., \& {Bleeker}, J.~A.~M. 1998, \aap,
  339, 201

\bibitem[{{Vink} \& {Laming}(2003)}]{vink03a}
{Vink}, J. \& {Laming}, J.~M. 2003, \apj, 584, 758

\bibitem[{{Warren} {et~al.}(2005)}]{warren05}
{Warren}, J.~S. {et~al.} 2005, \apj, 634, 376

\bibitem[{{Woosley} {et~al.}(2007){Woosley}, {Kasen}, {Blinnikov}, \&
  {Sorokina}}]{woosley07}
{Woosley}, S.~E., {Kasen}, D., {Blinnikov}, S., \& {Sorokina}, E. 2007, \apj,
  662, 487

\end{thebibliography}

\end{document}